\documentclass[11pt]{article}  
\usepackage{times,amsmath,amsfonts,amsthm,amssymb,eucal}
\usepackage{array}
\usepackage{color}
\usepackage{enumerate}
\usepackage{epsfig}
\usepackage{mathrsfs}
\usepackage{fancyhdr}
\usepackage{graphicx}
\usepackage{caption}
\usepackage{subcaption}
\usepackage{graphicx}
\usepackage{makeidx}
\usepackage{multirow}
\newcommand{\E}{\mathrm{E}}
\newcommand{\Expect}{{\rm I\kern-.3em E}}
\newtheorem{theorem}{Theorem}

\newtheorem{proposition}{Proposition}
\newtheorem{lemma}{Lemma}
\newtheorem{corollary}{Corollary}
\newtheorem{definition}{Definition}
\newtheorem{assumption}{Assumption}
\def\Theorem{\begin{theorem}\sl}
\def\EndTheorem{\end{theorem}}
\def\Proposition{\begin{proposition}\sl}
\def\EndProposition{\end{proposition}}
\def\Lemma{\begin{lemma}\sl}
\def\EndLemma{\end{lemma}}
\def\Corollary{\begin{corollary}\sl}
\def\EndCorollary{\end{corollary}}
\def\Definition{\begin{definition}\sl}
\def\EndDefinition{\end{definition}}
\numberwithin{equation}{section}

\begin{document}
\title{ \textbf{Bootstrapping the  Mean Vector for the Observations in the Domain of Attraction of a Multivariate Stable Law}}   
\author{Maryam Sohrabi and Mahmoud Zarepour\\ University of Ottawa, Ontario, Canada}
\date{\today}
\maketitle
\begin{abstract}
We consider a robust estimator of the mean vector for a sequence of independent and identically distributed (i.i.d.) observations in the domain of attraction of a stable law with different indices of stability, $DS(\alpha_1, \ldots, \alpha_p)$, such that $1<\alpha_{i}\leq 2$, $i=1,\ldots,p$.  The suggested estimator is asymptotically Gaussian with unknown parameters. We apply an asymptotically valid  bootstrap to construct a confidence region for the mean vector. A simulation study is performed to show that the estimation method  is efficient   for conducting inference about the mean vector for multivariate heavy-tailed distributions.

\bigskip
\noindent\textsc{Key words}:  Infinite Variance, Estimation, Resampling, Mean.
\end{abstract}

\pagestyle {myheadings} \markboth {} {Bootsrraping the mean vector}

{\it MSC 2010 Subject Classification}: Primary 62H10; 62G20; Secondary 62G09; 62G35.

\section {Introduction}
\label{IntM}

Let $X_1,X_2,\ldots,X_n$  be a sequence of i.i.d. random variables  from some distribution  $F$ with mean $\mu$.
Traditionally,  studentization has been  considered to make inference about the mean for relatively light-tailed distributions.  This approach requires a finite second moment, which is often not the case for a heavy-tailed law. Moreover, bootstrap inference is  arguably  accurate and is a simple approach to make inference for the univariate mean of finite variance observations; see Diciccio and Efron (1996) and Singh (1981).

Now suppose that $\{X_k\}$ are in the domain of attraction of a stable law with infinite second moment. In other words, there exist  constants $\{a_n>0\}$ and $\{b_n\}$   such that
$$S_n=a_n^{-1}\sum_{k=1}^n(X_k-b_n)\overset{d}\to S_\alpha,$$
where $S_\alpha$ is a stable random variable with index $0<\alpha\leq2$.
 It is known that $a_{n}=n^{1/\alpha}L(n)$,  where  $L$ is a  slowly varying  functions  at $\infty$; see Feller (1971) for more details. Throughout this paper, we assume that  $1<\alpha\leq2$  and $\E(X_1)<\infty$, so we can take $b_n=\mu$. Since the sample mean $\bar{X}_n$ is the usual estimator of the mean, it is natural to base inference about $\mu$ on $\bar{X}_n$.  Despite the fact that the sample mean is an intuitive estimate for the population mean, the rate of convergence of the sample mean is $na_n^{-1}$ which approaches to zero very slowly when $\alpha$ is near 1.

In a computational settings, properties of the various bootstrap procedures for the mean of heavy-tailed distributions have been considered extensively in the statistical literature; see, for example,
  Hall (1990) and Knight (1989a).  It has been shown that the regular bootstrap is not consistent for estimating the distribution of the mean. For finite variance observations, the bootstrap distribution of the sample mean converges almost surely to a fixed distribution. On the other hand, Athreya (1987) shows that the bootstrap distribution of the sample mean of infinite variance observations converges in distribution to a random probability
distribution.  Athreya, Lahiri, and Wu (1998) demonstrate that bootstrapping based on $m$ out of $n$ resampling, such that $m/n\to 0$, rectifies the asymptotic failure of the regular bootstrap for   heavy-tailed distributions.  They also consider the bootstrap methods for conducting inference about the mean of a sequence of i.i.d. random variables in the domain of attraction of a stable law whose index exceeds 1. Arcones and Gin\'{e} (1989) discuss almost sure and in probability bootstrap central limit theorem when the random variable $X$ is in the domain of attraction of  a stable law with infinite second moment.  Hall and LePage (1996)  propose a bootstrap method for estimating the distribution
of the studentized mean under  more general conditions on  the tails of the sampling distributions. They also show that this method holds even when  the sampling distribution is not in the domain of attraction of any limit law. Zarepour and Knight (1999b)   consider the weak limit behavior of a point process obtained by replacing the original observations by the bootstrap sample.

In this paper, we wish to make inference about the mean vector $\boldsymbol{\mu}$ of a multivariate heavy-tailed distribution. Consider the model
\begin{align}
\label{1.1}
\mathbf{X}_i=\boldsymbol{\mu}+\boldsymbol{\epsilon}_{i},
\end{align}
where  ${\mathbf{X}_i}=\left(X_{i1},\ldots,X_{ip}\right)$, $i=1,\ldots,n$,  are $\mathbb{R}^p$-valued random vectors, and  $\boldsymbol{\mu}=\left(\mu_1, \ldots , \mu_p\right)$ is an unknown fixed parameter vector. Let $\{\boldsymbol{\epsilon} _{i}\}=\{\left(\epsilon_{i1},\ldots ,\epsilon_{ip}\right)\}$ form a sequences of i.i.d. random vectors with  zero mean in the domain of attraction of a multivariate stable law. The traditional definition of the domain of attraction of a multivariate stable  law assumes that indices of stability are the same; see Rvaceva (1962).  However, for the multivariate case, observations can be in the domain of attraction of a stable law with different indices of stability. In many real life examples, some coordinates may have lighter tails while the other coordinates may have heavier tails. In this paper, we assume that errors are in $DS(\alpha_1,\dots,\alpha_p )$ with possibly different values of $\alpha_j \in (1,2]$  for $j=1,\ldots,p$. The definition of  the multivariate stable law with possibly different indices of stability  provided in  Resnick and Greenwood (1979) is as follows.

\begin{definition}
\label{DAD}
Given $\{ \mathbf{X}_i=\left(X_{i1} , X_{i2} , \ldots , X_{ip}\right)\}$ i.i.d. random vectors on $\mathbb{R}^p$ with distribution $\mathrm{F}$, let $\mathbf{S}_n = \sum_{i=1}^n \mathbf{X}_{i}$ and $S_n^{(j)} = \sum_{i=1}^nX_{ij}$ for $j=1,\ldots,p$. Then, $\mathrm{F}\in DS(\alpha_1,\ldots,\alpha_p)$, $\alpha_j\in(0,2]$, if there exist sequences $\mathbf{a}_n=(a_n^{(1)},\ldots,a_n^{(p)}),\mathbf{b}_n\in \mathbb{R}^p$ with $a_n^{(j)} > 0$  such that
\begin{align}
\label{JCAF}
\left(S_{n}^{(1)}/a_n^{(1)},\ldots,S_{n}^{(p)}/a_n^{(p)}\right)-\mathbf{b}_n\overset{d}\to \mathbf{Y},
\end{align}
where $\mathbf{Y}$ is a  random vector on $\mathbb{R}^p$
with stable distribution.
\end{definition}
The limiting distribution in \eqref{JCAF} can be a combination of stable laws with different tail indices.  When $\alpha_j=2$, $j=1,\ldots,p$, then $\mathbf{Y}$ has a multivariate normal distribution.  When one or more indices is equal to 2, the Gaussian limits will be always independent from the limiting components with indices less than 2. For more discussion about the class of all possible limits in \eqref{JCAF} see Resnick and Greenwood (1979). Notice that Definition \ref{DAD} is not the standard definition of the domain of attraction of a multivariate stable law since it allows different indices of stability for each coordinate. In fact, the common definition of  the domain of attraction of a stable law   is a special case of  Definition \ref{DAD} when $\alpha_1=\alpha_2=\cdots=\alpha_p$.

It is obvious that  the limiting distribution of the sample mean depends on the tail indices when the errors are in the domain of attraction of a stable law. Thus, it is hard to derive any inference for the mean vector $\boldsymbol{\mu}$ based on the limit, especially when the limiting distributions of the coordinates may have different indices of stability. A bootstrap procedure may circumvent this difficulty
but, as mentioned before,  the ordinary bootstrap fails in this case.  Using an $m$ out of $n$ bootstrap, when $m/n\to0$,  typically resolves the problem.
Note that the choice of $m$ is a key point  and controversial.  See Bickel and Sakov (2008) for more details.

To rectify both the near inconsistency and the bootstrap failure of the least square estimates of $\boldsymbol{\mu}$, we suggest a robust estimate. In our approach, the proposed robust estimation method for the mean vector has higher rate of convergence compared to the sample mean.
Moreover, we prove that the regular bootstrap is applicable as the limiting distribution for the M-estimate is a multivariate normal distribution.

This paper is organized as follows. Section \ref{MEMean} presents our main theorem, robust estimation of the mean vector for a sequence of  i.i.d. observations in the domain of attraction of a stable law with different indices of stability, $DS(\alpha_1, \ldots, \alpha_p)$, such that $1<\alpha_{i}\leq 2$, $i=1,\ldots,p$.  The bootstrap procedure is discussed in Section \ref{BootMean}.
Section \ref{SimMean} presents some simulations supporting the  results of this paper.

\section {M-estimates  of the mean vector}
\label{MEMean}
Let vector $\boldsymbol{\mu}$ be the parameter  of interest and let $\mathbf{X}_1,\ldots,\mathbf{X}_n$ be a random samples satisfying \eqref{1.1}. The classical M-estimate for $\boldsymbol{\mu}$, denoted by $\hat{\boldsymbol{\mu}}_M$, is defined as the minimizer of the function
\begin{align}
\label{14}
\underset{\boldsymbol{\beta}}{\text{arg min}}\sum_{i=1}^{n}\rho\left(\mathbf{X}_i-\boldsymbol{\beta}\right),
\end{align}
where $\rho$ is an almost everywhere differentiable convex function. This guarantees the uniqueness of the solution. For more details see Davis, Knight, and Liu (1992).  Note that the parameter estimate $\hat{\boldsymbol{\mu}}_M$ minimizing the objective function  \eqref{14} also minimizes the modified objective function
\begin{align*}
 \sum_{i=1}^n\left[\rho(\mathbf{X}_i-\boldsymbol{\beta})-\rho(\epsilon_t)\right],
 \end{align*}
 which can be rewritten as
 \begin{align*}
   \sum_{i=1}^n\left[\rho\left(\boldsymbol{\epsilon_t}-n^{1/2}(\boldsymbol{\beta}-\boldsymbol{\mu})n^{-1/2}\right)-\rho(\epsilon_t)\right].
 \end{align*}
 For convenience, similar to Zarepour and Roknossadati (2008), we consider the multivariate loss function as
\begin{align}
\label{MultiHub}
{\rho}\left( x_1 ,\ldots,x_p \right)= {\rho _1}\left( x_1 \right)+\cdots+ {\rho _p}\left( x_p \right),
\end{align}
where ${\rho _j}(\cdot)$, $j=1,\ldots,p$, are univariate loss functions. A good justification for using the objective function of the form \eqref{MultiHub} is the ability to calibrate with respect to the thickness of the tails for  each coordinate to derive more precise estimates in  practice.

Here, we impose the following assumptions on the functions $\rho_j$,  for $j=1,\ldots,p$.
\begin{assumption}
(A1) $\rho_j(\cdot):\mathbb{R}\rightarrow\mathbb{R}$ is a convex and twice differentiable function, and take $\psi_j(\cdot)=\rho'_j(\cdot)$, and    $\psi'_{j}(\cdot)=\rho''_j(\cdot)$.
\end{assumption}
  \begin{assumption} (A2)
$\E(\psi_j(\epsilon_{1j}))=0$, $\E(\psi_j^{2}(\epsilon_{1j}))<\infty$, and $0<|\E(\psi'_{j}(\epsilon_{1j}))|<\infty$.
\end{assumption}
\begin{assumption} (A3)
  $\psi_{j}(\cdot)$ has Lipschitz-continuous derivative $\psi'_{j}(\cdot)$; i.e., there exists a real constant $k\geqslant 0$ such that for all $x$ and $y$,
     \begin{align*}
      |\psi_j^{'}(x)-\psi_j^{'}(y)|\leq k|x-y|.
     \end{align*}
\end{assumption}
\noindent Assumptions A1-A3 are standard when dealing with the asymptotic behaviour of M-estimates. The only real restriction is the assumption that  $\psi_j(\cdot)$ may not be differentiable everywhere.
However, in this case the results will usually hold with some additional complexity in the proofs; for more details see Knight (1989b).
\begin{definition}
Let $(\Omega,\mathcal{A},P)$ be a probability space and $(a,b)\in\mathbb{R}$ be an interval. We say that a stochastic process $T:(a,b)\times\Omega\to\mathbb{R}$ is convex if
$$T(\lambda s+(1-\lambda)t,\cdot)\leq\lambda T(s,\cdot)+(1-\lambda) T(t,\cdot)$$
almost everywhere for all $s,t\in(a,b)$ and $\lambda\in[0,1]$.
\end{definition}
\noindent The following lemma  is used to prove our main results.

\begin{lemma}
\label{Asy.Tn3}
Suppose that  $\{T_n(\cdot)\}$ is a sequence of convex stochastic processes  on $\mathbb{R}$ and suppose that
$$T_n(\cdot)\overset{d}\to T(\cdot).$$
 Then $\{T_n(\cdot)\}$ has  a unique minimum $\boldsymbol{\kappa}_n$.  If $\boldsymbol{\kappa}$ minimizes $T(\cdot)$, then
$$\boldsymbol{\kappa}_n\overset{d}\to\boldsymbol{\kappa}.$$
\end{lemma}
\noindent {\textbf{Proof.}} The proof is given in Lemma 2.2 of Davis et al. (1992). See also Knight (1989b).

$\hfill\square$
\\
\begin{theorem}
\label{CONF INT}
Suppose \eqref{1.1} holds. With the loss function defined in \eqref{MultiHub}, let  $\hat{\boldsymbol{\mu}}_M$ be the M-estimator of the mean vector for a sequence of  i.i.d. observations in the domain of attraction of a stable law with indices of stability $(\alpha_1, \ldots, \alpha_p)$ such that $1<\alpha_{j}\leq 2$, $j=1,\ldots,p$. Then,  we have
\begin{align}
\label{MUlim}
\mathbf{W}_n=\sqrt{n}(\hat{\boldsymbol{\mu}}_M-\boldsymbol{\mu})\overset{d}\to \mathbf{W}.
\end{align}
Here, $ \mathbf{W}=\left(W_1, \ldots , W_p\right)^T\intercal$  has a multivariate normal distribution with zero mean vector and the covariance matrix
$\Sigma=\digamma^{-1}\Gamma\digamma^{-1}$ where
\begin{align}
\label{Cov1}
\digamma=\rm{diag}\left(\E(\psi_1^{'}(\epsilon_{11})),\ldots,\E(\psi_p^{'}(\epsilon_{1p}))\right)
\end{align}
 and  $\Gamma=(\gamma_{jk})$ is a $p \times p$ matrix such that
\begin{align}
\label{Cov2}
\gamma_{jk}={\rm Cov}\left(\psi_j(\epsilon_{1j}),\psi_k(\epsilon_{1k})\right)=\E\left(\psi_j(\epsilon_{1j})\psi_k(\epsilon_{1k})\right)
\end{align}
 for  $j,k=1,\ldots,p$.
\end{theorem}
\vspace{1cm}
\noindent {\textbf{Proof:}}
Under conditions A1-A3, define the convex process
\begin{align}
\label{z1}
\nonumber A_{n}(\mathbf{u})&= \sum_{j=1}^p\sum_{i=1}^n\left(\rho_j\left(\epsilon_{ij}-n^{-1/2}u_j\right)-\rho_j(\epsilon_{ij})\right)\\
&=\frac{-1}{\sqrt{n}}\sum_{j=1}^pu_j\sum_{i=1}^{n}\psi_j(\epsilon_{ij})+\frac{1}{2n}\sum_{j=1}^pu_j^{2}\sum_{i=1}^{n}\psi_j^{'}(c_{ij}),
\end{align}
where $\mathbf{u}^T=(u_1,\ldots,u_p)$ and $u_j=n^{1/2}(\hat{\mu}_{Mj}-\mu_j)$, for $j=1,\ldots,p$. Note that $c_{ij}$ is between $\epsilon_{ij}$ and $\epsilon_{ij}-n^{-1/2}u_j$. Using the Lipschitz continuity of $\psi'_j(\cdot)$, A3, we get
\begin{align}
\label{LC}
n^{-1}\sum_{j=1}^{p}|\psi'_j(\epsilon_{ij})-\psi'_j(c_{ij})|
\le k n^{-1}\sum_{j=1}^{p}|n^{-1/2}u_j|  \overset{P}\to  0.
\end{align}
Therefore, \eqref{LC} implies that $\psi'_j(c_{ij})$ can asymptotically be replaced by  $\psi'_j(\epsilon_{ij})$ in \eqref{z1}. It is also well known that
\begin{align}
\label{ZDif}
\frac{1}{\sqrt{n}}\left(
\begin{array}{ccc}
 \sum_{i=1}^{n}\psi_1(\epsilon_{i1})\\
 \vdots\\
\sum_{i=1}^{n}\psi_p(\epsilon_{ip})
 \end{array}
\right)\overset{d}\to \mathbf{Z},
\end{align}
where $\mathbf{Z}=\left(Z_1,\ldots,Z_p\right)^T$ has a multivariate normal distribution with zero mean vector and covariance matrix $\Gamma$ defined by \eqref{Cov2}. Note that   A2 implies $\gamma_{jk}<\infty$,  for  $j,k=1,\ldots,p$. Therefore,
$$A_{n}(\mathbf{u})\overset{d}\to A(\mathbf{u})=-\mathbf{u}^T\mathbf{Z}+
\frac12\mathbf{u}^T\digamma\mathbf{u},$$
where  $\digamma$ and $\mathbf{Z}$ are defined in \eqref{Cov1} and \eqref{ZDif}, respectively.
 From Lemma \ref{Asy.Tn3}, the minimizer of $A_{n}$ which is
$n^{1/2}(\hat{\boldsymbol{\mu}}_{M}-\boldsymbol{\mu})$ converges to the minimizer of $A$. Thus,
 \eqref{MUlim} follows by setting the derivative of $A(\mathbf{u})$ to 0 and solving for $(u_1,u_2,\dots,u_p)$.

$\hfill\square$
\\

Based on Theorem \ref{CONF INT}, a simple approach to construct a $100(1-\alpha)\%$  confidence region  for the mean of a $p-$dimensional  random vector in the domain of attraction of a multivariate stable law with large sample size is the ellipsoid determined by all  $\boldsymbol{\mu}$ such that
\begin{align}
\label{ConfRG}
CR_{1-\alpha}=\{\boldsymbol{\mu}: n(\hat{\boldsymbol{\mu}}_M-\boldsymbol{\mu})^T\mathbf{S}^{-1}(\hat{\boldsymbol{\mu}}_M-\boldsymbol{\mu})\leq \tau_{1-\alpha}\}.
\end{align}
Here and this context throughout, $\tau_{1-\alpha}$ is the $1-\alpha$ quantile of a $\chi^2(p)$ distribution and  $\mathbf{S}$ is the estimated value of $\Sigma$ by replacing $\boldsymbol{\epsilon}_{i}$ with the residuals,  $\boldsymbol{e}_{i}=\mathbf{X}_i-\hat{\boldsymbol{\mu}}_M $, $i=1,2,\ldots,n$, in  \eqref{Cov1} and \eqref{Cov2}.

 The confidence region in \eqref{ConfRG} gives the joint knowledge concerning reasonable values of $\boldsymbol{\mu}$ when the correlation between the measured variables is taken into account. Typically, any summary of conclusions includes confidence statements about the individual component means. Let  \eqref{1.1} hold for i.i.d. of $p$-dimensional random vectors $\mathbf{X}_1,\mathbf{X}_2,\ldots,\mathbf{X}_n$. Consider the following linear combination
 $$\mathbf{a}^T\mathbf{X}_i=a_1{X}_{i1}+a_2{X}_{i2}+\ldots+a_p{X}_{ip},\ \ i=1,2,\ldots,n.$$
For all $\mathbf{a}$, the interval
 $$\left(\mathbf{a}^T\hat{\boldsymbol{\mu}}_M-\sqrt{\tau_{1-\alpha}\mathbf{a}^T\mathbf{S}\mathbf{a}/n}\ ,\ \mathbf{a}^T\hat{\boldsymbol{\mu}}_M+\sqrt{\tau_{1-\alpha}\mathbf{a}^T\mathbf{S}\mathbf{a}/n}\right)$$
 contains $\mathbf{a}^T\hat{\boldsymbol{\mu}}_M$ with probability $1-\alpha$. The consecutive choices $\mathbf{a}^T=(1,0,\ldots,0)$, $\mathbf{a}^T=(0,1,\ldots,0)$, and so on through $\mathbf{a}^T=(0,0,\ldots,1)$ for the $\chi^2-$intervals allow us to conclude that
 \begin{align*}
&P(\hat{\mu}_{M1}-\sqrt{\tau_{1-\alpha}s_{11}/n}\leq \mu_1 \leq \hat{\mu}_{M1}+\sqrt{\tau_{1-\alpha}s_{11}/n})\\
 &=P(\hat{\mu}_{M2}-\sqrt{\tau_{1-\alpha}s_{22}/n}\leq \mu_2 \leq \hat{\mu}_{M2}+\sqrt{\tau_{1-\alpha}s_{22}/n})\\
 &=\cdots\\
&=P(\hat{\mu}_{Mp}-\sqrt{\tau_{1-\alpha}s_{pp}/n}\leq \mu_p \leq \hat{\mu}_{Mp}+\sqrt{\tau_{1-\alpha}s_{pp}/n})\\
&=1-\alpha.
 \end{align*}

\section {Bootstrapping the Mean Vector}
\label{BootMean}
It has been pointed out that the regular bootstrap fails to estimate the distribution of the sample mean of heavy-tailed observations. The main reason
for the failure of the regular bootstrap comes from the fact that  rare events occur when we resample the data. This means the resampling procedure will remember the magnitude of the observations in the resampled data. This fact is reflected in point process theory which is used as a tool for the asymptotic analysis of heavy-tailed observations.
Let  $\mathbf{X}_1,\mathbf{X}_2,\ldots$ be a sequence of i.i.d. random vectors in $DS(\alpha_1,\ldots,\alpha_p)$. Given $\mathbf{X}_1,\ldots,\mathbf{X}_n$, we draw an i.i.d. sequence of observations $\mathbf{X}_1^*,\ldots,\mathbf{X}_n^*$ from the empirical distribution
$$F_n(\cdot)=\frac1 n\sum _{i=1}^{n}\varepsilon_{\mathbf{X}_i}(\cdot),$$
where $\varepsilon_x$ is  defined by $\varepsilon_x(\cdot)=I(x \in \cdot)$ and $I$ is the indicator function. Define $M_{i,n}^*=\sum_{k=1}^nI(\mathbf{X}_k^*=\mathbf{X}_i)$. By Lemma 3.1 of Zarepour and Knight (1989a), we have
$$\left(M_{1,n}^*,\ldots,M_{n,n}^*,0,0,\ldots\right)\overset{d}\to\left(M_{1}^*,\ldots,M_{n}^*,\ldots\right),$$
where $M_1^*,M_2^*,\ldots$ are i.i.d. Poisson(1) random variables. Let $\mathbf{X}_i^*=\left(X_{i1}^* ,\ldots , X_{ip}^*\right)$. Now, for the corresponding sequence of point processes, we have
$$\sum_{i=1}^n\varepsilon_{\left((a_n^{(1)})^{-1}X_{i1}^*,\ldots,(a_n^{(p)})^{-1}X_{ip}^*\right)}=\sum_{i=1}^nM_{i,n}^*\varepsilon_{\left((a_n^{(1)})^{-1}X_{i1},\ldots,(a_n^{(p)})^{-1}X_{ip}\right)}.$$
With  considerable help from Theorem 4 of  Resnick and Greenwood (1979) and Resnick (2004) and Zarepour and Knight (1999b), it can be shown that
{\small
\begin{align}
\label{PMC}
\sum_{i=1}^n\varepsilon_{\left((a_n^{(1)})^{-1}X_{i1}^*,\ldots,(a_n^{(p)})^{-1}X_{ip}^*\right)}\overset{d}\to
\sum_{i=1}^\infty M_i^*\varepsilon_{\left(\rm{sign}(\gamma_{i1})|\gamma_{i1}|^{1/\alpha_1}\Gamma_i^{-1/\alpha_1},\ldots,\rm{sign}(\gamma_{ip})|\gamma_{ip}|^{1/\alpha_p}\Gamma_i^{-1/\alpha_p}\right)},
\end{align}}
in distribution. Here, $\{\Gamma_1,\Gamma_2,\ldots\}$ is a sequence of arrival times of a Poisson process with unit arrival rate and $\boldsymbol{\gamma}_i=(\gamma_{i1},\ldots,\gamma_{ip})\sim G$ and $G$ is a distribution on the  boundary of unit sphere.   To have a valid bootstrap, we expect to have $M_i^*=1$ which is not the case here and $M_i^*$ are random quantities.

The limiting distribution for the bootstrap sample mean, $\bar{\mathbf{X}}^*$, can be derived from  \eqref{PMC} and the continuous mapping theorem along with extra mathematical steps. Similar to the univariate case,  this result shows that the regular bootstrap  fails asymptotically. As discussed in the introduction, a subsampling scheme ($m$ out of $n$ bootstrap such that $m/n\to 0$) is an appropriate approach
to achieve asymptotic validity of a bootstrap procedure for constructing a confidence region for the mean vector of
i.i.d. heavy-tailed data.  However, choosing the proper subsample size $m$ is of great concern and controversial to many authors.

On the other hand, Theorem \ref{CONF INT} shows that the weak limit behavior of $\sqrt{n}(\hat{\boldsymbol{\mu}}_M-\boldsymbol{\mu})$ is a multivariate normal distribution. Thus, the regular bootstrap works if we use the robust estimates (M-estimates) for the mean vector. Our approach in this section is to consider a bootstrap approach to estimate the confidence region for $\boldsymbol{\mu}$. Given $\mathbf{X}=\left(\mathbf{X}_1 ,\ldots , \mathbf{X}_n\right)$,  find M-estimates of $\boldsymbol{\mu}$ in model \eqref{1.1} using the objective function in \eqref{MultiHub}. Then calculate the residuals, where
\begin{align}
\label{1.2}
\boldsymbol{e}_{i}=\mathbf{X}_i-\hat{\boldsymbol{\mu}},\ \text{ }i=1,2,\ldots,n.
\end{align}
Let $\boldsymbol{e}_1^*,\ldots,\boldsymbol{e}_n^*$ be  a sample of size $n$ from the centered residuals in \eqref{1.2}. These assumptions imply the following lemma.
\begin{lemma}
 \label{SRS}
 Let $\{\boldsymbol{e}_1^*,\ldots,\boldsymbol{e}_n^*\}$ be an i.i.d. sample from $F_n(\cdot)=\frac{1}{n}\sum_{i=1}^{n}\varepsilon_{(\boldsymbol{e}_{i}-\bar{\boldsymbol{e}}\leq\cdot)}$  where $\boldsymbol{e}_i^*=\left(e_{i1}^* , \dots , e_{ip}^*\right)$, $i=1,\ldots,n$, and $\E^*$ denotes the expectation under $F_n$. Also,  let $\psi_j$, $j=1,\ldots,p$, satisfy conditions A2-A3. Then, for $j=1,\ldots,p$, we have
\begin{enumerate}[(i)]
\item $\E^*(\psi_j(e_{1j}^*))=0.$
\item Almost sure, we have
\begin{align*}\frac{1}{\sqrt{n}}\left(
\begin{array}{ccc}
 \sum_{i=1}^{n}\psi_1(e_{i1}^*)\\
 \vdots\\
\sum_{i=1}^{n}\psi_p(e_{ip}^*)
 \end{array}
\right)\overset{d}\to \mathbf{Z},
\end{align*}
where $\mathbf{Z}=\left(Z_1,\ldots,Z_p\right)^T$ has a multivariate normal distribution with zero mean vector and covariance matrix $\Gamma$ defined by \eqref{Cov2}.
\item $\E^*(\psi_j^{'}(e_{1j}^*))=\frac1n\sum_{i=1}^{n}\psi_j^{'}(e_{ij}^*)\overset{p}\to \E(\psi_j^{'}(\epsilon_{1j}))$.
\end{enumerate}
\end{lemma}
\vspace{5mm}
\noindent {\textbf{Proof:}}
 We omit the proofs here since the results follow by arguments similar to those of  Singh (1981) and  Moreno and Romo (2012).
$\hfill\square$
\\

\noindent Now, we are ready to  derive the limiting distribution of bootstrap estimates. Recall that  $\boldsymbol{e}_1^*,\ldots,\boldsymbol{e}_n^*$ are  a sample of size $n$ from the centered residuals in \eqref{1.2}. We calculate  $\mathbf{X}_i^*$, $i=1,\ldots,n$, from  $\mathbf{X}_{i}^*=\hat{\boldsymbol{\mu}}+\boldsymbol{e}_{i}^*$. Then, we minimize
\begin{align*}
A_{n}^*(u_1^*,\ldots,u_p^*)&= \sum_{j=1}^p\sum_{i=1}^n\left(\rho_j\left(e^*_{ij}-n^{-1/2}u_j^*\right)-\rho_j(e^*_{ij})\right)\\
&=\frac{-1}{\sqrt{n}}\sum_{j=1}^pu_j^*\sum_{i=1}^{n}\psi_j(e^*_{ij})+\frac{1}{2n}\sum_{j=1}^p{u_j^*}^{2}\sum_{i=1}^{n}\psi_j^{'}(c_{ij}^*),
\end{align*}
to get the minimizers $u_j^*=n^{1/2}(\hat{\mu}_{Mj}^*-\hat{\mu}_{Mj})$ for $j=1,\ldots,p$ and $c_{ij}^*$ is between $e_{ij}^*$ and $e_{ij}^*-n^{-1/2}u_j^*$. Then from Lemma \ref{SRS}, similar to Theorem 1, we get
\begin{align}
\label{BootConv}
\mathbf{W}_n^*=\sqrt{n}(\hat{\boldsymbol{\mu}}_M^*-\hat{\boldsymbol{\mu}}_M) \overset{d} \to \mathbf{W},
\end{align}
for almost every sample path, where $\mathbf{W}$ is defined in \eqref{MUlim}.  We carry out a large number, say $B$, of the bootstrap replicates of size $n$ from
$$C^*=n(\hat{\boldsymbol{\mu}}_M^*-\hat{\boldsymbol{\mu}}_M)^T\mathbf{S}^{*^{-1}}(\hat{\boldsymbol{\mu}}_M^*-\hat{\boldsymbol{\mu}}_M),$$
where $\mathbf{S}^*$ is the estimated value of $\Sigma$ by replacing $\boldsymbol{\epsilon}_i$ with the  bootstrap residuals  $\boldsymbol{e}_{i}^*=\mathbf{X}_i^*-\hat{\boldsymbol{\mu}}_M^* $, $i=1,2,\ldots,n$, in  \eqref{Cov1} and \eqref{Cov2}.
Set $\hat{\tau}_\alpha$ to be the $100\alpha$-th percentile value of $\{C^*(b),b=1,2,\ldots,B\}$.
Thus an approximate naive confidence region for $\boldsymbol{\mu}$ at level $100(1-\alpha)\%$ will be
\begin{align}
\label{ConfRGB}
 n(\hat{\boldsymbol{\mu}}_M-\boldsymbol{\mu})^T\mathbf{S}^{-1}(\hat{\boldsymbol{\mu}}_M-\boldsymbol{\mu})\leq \hat{\tau}_{1-\alpha}.
\end{align}

\section{Simulation}
\label{SimMean}
To illustrate the preceding results, some simulation  studies are performed. Our simulation takes the follow two steps. First, we consider how to choose a proper loss function and its tuning constants to estimate the population mean, $\boldsymbol{\mu}$.  In this step, we also conduct the numerical studies
to compare the performance of the M-estimate to the traditional least square estimate, $\bar{\mathbf{X}}$, which is considered by    Athreya et al. (1998). Notice that, for the sake of simplicity, we only consider the univariate case  for the comparison purposes in our simulation study. This is due to the fact that the asymptotic distribution of the sample mean for the multivariate observations in the domain of attraction of a stable law with different indices of stability has not been considered in statistical literature and is not easily computationally tractable. However, in the second step, we also consider the asymptotic results of M-estimates for  a bivariate case when errors are in the domain of attraction of a stable law with different indices of stability. Our simulation scheme is as follows.

\begin{figure}[h!]
  \centering
  \includegraphics[width=0.53\textwidth]{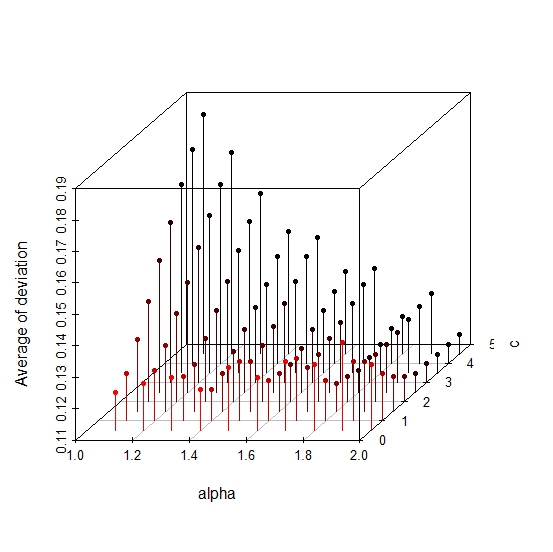}
 \caption{Average values of $|\hat{\mu}_M-\mu|$ for different values of $\alpha$ and truncation values of $c$ in the Huber loss function with the replication size of 10,000.}
         \label{fig:AD}
\end{figure}

\bigskip
\noindent
{\bf Step 1}: The first step in our simulation study is choosing the loss function. An example for the univariate $\rho_{j}(\cdot)$ is the Huber  loss function given by
\begin{align}
\label{HLF}
{\rho_{c_j}}\left( x \right)
= \left\{ {\begin{array}{*{20}{l}}
{\tfrac{1}{2}{x^2}}&{\rm{if}\,}\left| x \right| \le c_j,\\
{c_j\left| x \right| - {{\tfrac{1}{2}}{c_j^2}}}&{\rm{if}\,}\left| x \right| > c_j,
\end{array}} \right.
\end{align}
for a known constant $c_j$; see Huber (1981). Then,  $\psi_{c_j}(x)=\max[\min(x,c_j),-c_j]$. The choice of a truncation value  $c_j$ is of practical interest especially when we have different indices of stability. The following univariate simulation study is undertaken in order to explore whether there is a relationship between values of $c_j$ in  the Huber loss function and the index of stability $\alpha$. Consider the univariate model
 \begin{align}
 \label{US}
 X_{i}=\mu+\epsilon_{i}, \text{ }i=1,2,\ldots,n,
\end{align}
where $\{\epsilon_{i}\}\in DS(1<\alpha\leq 2)$. We generate the random samples $\{X_i\}_{i=1}^n$ in model \eqref{US} for $\mu=3$ and $n=100$ with different values of $1<\alpha\leq 2$. Then,  the M-estimates of $\mu$ in \eqref{US} are calculated from the generated random samples using the Huber loss function given in \eqref{HLF}. To seek a more efficient $c$ in  the Huber loss function,  the estimation is repeated for different values of $c$ between $0.5$ and $4.5$ for each choice of $\alpha$.  We  find the average deviation by calculating the absolute deviation $|\hat{\mu}_{M}-\mu|$ and then carrying out  10,000 replications. The numbers in Table \ref{tab:TAlphaC} are  the averages of the replications. Meanwhile, the scatterplot of the  average deviations is presented in Figure \ref{fig:AD}.
For each level of $\alpha$ and $c$, the minimum of the average  deviations appear in boldface in Table \ref{tab:TAlphaC}. This table shows that, for instance, for $1.1\leq\alpha\leq1.4$ if we choose $c=1$, we get the minimum of the error estimation.
 Table \ref{tab:TAlphaC} and Figure \ref{fig:AD} also show that there is a positive relationship between the truncation value $c$ and the index of stability $\alpha$. In fact, to have less estimation error, we must choose a larger value of $c$ as $\alpha$ gets larger and this conclusion is not surprising.
\begin{table}[h!]
\centering
\caption{Average values of $|\hat{\mu}_M-\mu|$ for different values of $\alpha$ and truncation values of $c$ in the Huber loss function with the replication size of 10,000.}
\begin{tabular}{lccccccccc}
\hline
  {} & {} &  {} & {} &  {} &  $c$ & {} &  {} & {} & {}\\\cline{2-10}
  $\alpha$ & 0.5 & 1 & 1.5 & 2 & 2.5 &  3 &  3.5 & 4 & 4.5 \\
  \hline
1.1 & \bf{0.123} & \bf{0.126} & 0.134 & 0.143 & 0.153 & 0.162 & 0.171 & 0.180 & 0.188\\
1.2 & \bf{0.125} & \bf{0.125} & 0.131 & 0.138 & 0.146 & 0.154 & 0.161 & 0.169 & 0.176\\
1.3 & 0.128 & \bf{0.126} & \bf{0.127} & 0.136 & 0.142 & 0.149 & 0.156 & 0.162 & 0.168\\
1.4 & 0.128 & \bf{0.125} & \bf{0.127} & 0.131 & 0.137 & 0.142 & 0.147 & 0.153 & 0.158\\
1.5 & 0.129 & \bf{0.125} & \bf{0.125} & \bf{0.128} & 0.132 & 0.136 & 0.140 & 0.145 & 0.148\\
1.6 & 0.130 & 0.128 & \bf{0.123} & \bf{0.125} & \bf{0.126} & 0.131 & 0.134 & 0.137 & 0.141\\
1.7 & 0.130 & 0.124 & \bf{0.122} &\bf{ 0.122} & \bf{0.123} & 0.125 & 0.127 & 0.130 & 0.132\\
1.8 & 0.129 & 0.122 & 0.120 & \bf{0.118} & \bf{0.118} & \bf{0.119} & 0.121 & 0.122 & 0.124\\
1.9 & 0.131 & 0.124 & 0.120 & \bf{0.118} & \bf{0.117} & \bf{0.118} & \bf{0.118} & 0.119 & 0.120\\
2.0 & 0.131 & 0.123 & 0.119 & \bf{0.116} & \bf{0.115} & \bf{0.114} & \bf{0.114} & \bf{0.114} & \bf{0.114}\\
\hline
\end{tabular}
\label{tab:TAlphaC}
\end{table}

In order to study in-depth the refinement provided by M-estimates
 over the classical counterpart $\bar{X}$, we first consider the
empirical probabilities of coverage of  these estimates for $\mu$ in Table \ref{tab:CPH2}. We generated samples of size 100 from Model \eqref{US} when  $\mu=1$ and $\{\epsilon_{i}\}\in DS(1<\alpha\leq 2)$ for $\alpha=1.1,1.3,1.5,1.8$, and $2.0$.
We are interested in estimating the population mean $\mu$. To compare the sample mean to M-estimate method,   we consider the following steps to calculate the  coverage probability for the two sided bootstrap confidence intervals.
\begin{enumerate}[(i)]
\item We estimate  the parameter $\mu$ in Model \eqref{US} using the least square method, $\bar{x}$, and the M-estimate method, $\hat{\mu}_M$. For the M-estimates, we apply the    Huber loss function with different values of $c$ from Table \ref{tab:TAlphaC}  according to different values of $\alpha$. Then, for each estimate,  we take
$$e_i=X_i-\hat{\mu}, \ i=1,\ldots,100.$$
\item Due to the fact that the regular bootstrap is not valid when we use the sample mean for heavy-tailed samples (Athreya et al. (1998)), we draw a sample of size $m$ from centered residuals denoted by  $e_1^*,\ldots,e_m^*$, and we find  $\{X_i^*\}_{i=0}^m$ from \eqref{US}. Then we  estimate the parameter $\mu$ by getting the average of the bootstrap samples. On the other hand, the regular bootstrap is valid when we apply M-estimate method. Therefore, by taking $m=n$, we estimate $\mu$ by M-estimate method
    using the bootstrap observations  by  the same minimization of the objective function used in  step (i).
\item  We repeat  step (ii)  for $B=2,000$ times to get $\bar{x}^{1*},\ldots,\bar{x}^{2,000*}$ and $\hat{\mu}_M^{1*},\ldots,\hat{\mu}_M^{2,000*}$.  To find a  naive $100(1-\alpha)\%$ confidence interval for $\mu$, we obtain the $100(\alpha/2)\%$-th and $100(1-\alpha/2)\%$-th percentiles  of   2,000 bootstrap estimates  as  the lower and upper bound of our confidence intervals.
\end{enumerate}

In order to compute the coverage probability of the bootstrap confidence intervals, the original  observations are generated 1,000 times for each choice of $\alpha$. Moreover, to study  how the selection of the resampling size would affect the estimation based on the sample mean, we perform the second step with three different resampling sizes $m=n/\ln(\ln(n))$, $n^{0.9}$,  and $n^{0.95}$. Notice that the choice of $m$ is not only  controversial  but also  is a necessary value for a consistent resampling if we insist on using the least square estimation.   Then by applying (i)-(iii), the naive  90\%, 95\%, and 99\% bootstrap confidence intervals for $\mu$  using the sample mean and the M-estimates are tabulated in Table \ref{tab:CPH2}. Table \ref{tab:CPH2} below shows that the coverage probabilities for the M-estimates  in each case,  for all values of $\alpha$, are close enough to  the nominal confidence levels. This table also indicates
a  robust performance of the M-estimates as compared to the sample mean.
\begin{table}[!h]
\small
\centering
\caption{Comparing the empirical probability coverage of the M-estimate  and the sample mean   for sample size $n=100$}
\begin{tabular}{lcccccc}
  \hline
   {} &  {} & \multirow{2}{*}{M-estimation} & \multicolumn{3}{c}{Sample mean} \\
  \cline{4-6}
  ${}$ &  {} &  & {$m=\frac{n}{ln(ln(n)}$} & {$m=n^{(0.9)}$} & {$m=n^{(0.95)}$} \\
  \hline
  \multirow{3}{*}{$\alpha=1.1$} & $I_{0.90}$ &  0.891 &  0.893 & 0.854 & 0.855  \\
  {} & $I_{0.95}$ &  0.937 & 0.938 & 0.959 & 0.972   \\
  {} & $I_{0.99}$ &  0.985 & 0.960 & 1.000 & 1.000\\
       \hline
  \multirow{3}{*}{$\alpha=1.3$} & $I_{0.90}$ &    0.891 & 0.841 & 0.834 & 0.840  \\
  {} & $I_{0.95}$ &  0.950 & 0.911 & 0.961 & 0.961\\
  {} & $I_{0.99}$ & 0.991 & 0.938 & 0.999 & 1.000 \\
       \hline
  \multirow{3}{*}{$\alpha=1.5$} & $I_{0.90}$ &  0.889 &  0.840 & 0.894 & 0.899   \\
  {} & $I_{0.95}$ &  0.936 & 0.874 & 0.967 & 0.971   \\
  {} & $I_{0.99}$ &   0.981 & 0.933 & 1.000 & 1.000\\
       \hline
    \multirow{3}{*}{$\alpha=1.8$} & $I_{0.90}$ &  0.889 & 0.813 & 0.944 & 0.943    \\
  {} & $I_{0.95}$ & 0.941 & 0.859 & 0.991 & 0.991  \\
  {} & $I_{0.99}$ &  0.986 & 0.934 & 0.999 & 0.998\\
   \hline
  \multirow{3}{*}{$\alpha=2$} & $I_{0.90}$ &  0.888 & 0.731 & 0.945 & 0.953    \\
  {} & $I_{0.95}$ & 0.949 & 0.830 & 0.999 & 0.999 \\
  {} & $I_{0.99}$ &   0.984 & 0.916 & 1.000 & 1.000\\
       \hline
 \end{tabular}
\label{tab:CPH2}
\end{table}

To compare the point estimates, we perform a
numerical illustrations of the mean   absolute  deviation of the estimates from $\mu$ in Table \ref{tab:CPH3}. The table  indicates a significantly better performance of the
M-estimates for $\mu$ over its classical counterpart $\bar{X}$, especially for $\alpha$ close to 1.
\begin{table}[!h]
\small
\centering
\caption{Mean  absolute deviation of  the estimates from $\mu$  for sample size $n=100$}
\begin{tabular}{lcccccc}
  \hline
     {} &    \multicolumn{5}{c}{$\alpha$} \\
  \cline{2-6}
  ${}$ &  {1.1} &  {1.3} & {1.5} & {1.8} & {2.0}\\
  \hline
  $|\hat{\mu}_M-1|$ & 0.1250 &  0.1262 &  0.1270 & 0.1170 & 0.1137  \\
  $|\bar{x}-1|$ &  1.9040 &    1.1357 & 0.5117  & 0.1530  & 0.1110   \\
  \hline
 \end{tabular}
\label{tab:CPH3}
\end{table}

\bigskip

\noindent{\bf Step 2}:  Now consider the bivariate model
 \begin{align}
 \label{BiM}
\left(
{\begin{array}{*{20}{c}}
X_{i1}\\
X_{i2}
\end{array}}
\right)=\left(
{\begin{array}{*{20}{c}}
\mu_{1}\\
\mu_{2}
\end{array}}
\right)+\left(
{\begin{array}{*{20}{c}}
\epsilon_{i1}\\
\epsilon_{i2}
\end{array}}
\right), \text{ }i=1,2,\ldots,n.
\end{align}
Set $\left(\mu_1 , \mu_2\right)=\left(1 , 14\right)$ and  $\{\boldsymbol{\epsilon}_{i}\}$ are in a domain of attraction of a symmetric bivariate stable laws with indices of stability $1<\alpha_1\leq2$ and $1<\alpha_2\leq2$. To generate  $\{\boldsymbol{\epsilon}_1\}=\{\left(\epsilon_{11} , \epsilon_{12}\right)\}$ in \eqref{BiM} with the preceding indices of stabilities, consider the set of $K=10,000$ points $\{\boldsymbol{\gamma}_i=(\cos\theta_i,\sin\theta_i):\theta_i\in[0,2\pi],\ i=1,\ldots,10000\}$  on the boundary of the unit circle.
By \eqref{PMC}, we draw the error $\{\boldsymbol{\epsilon}_1\}$  from
\begin{align}
\label{ME}
\boldsymbol{\epsilon}_1=\left(\epsilon_{11}, \epsilon_{12}\right)=\left(\sum_{i=1}^K\rm{sign}(\boldsymbol{\gamma}_{i1})|\boldsymbol{\gamma}_{i1}|^{1/\alpha_1}\Gamma_i^{-1/\alpha_1}
, \sum_{i=1}^K\rm{sign}(\boldsymbol{\gamma}_{i2})|\boldsymbol{\gamma}_{i2}|^{1/\alpha_2}\Gamma_i^{-1/\alpha_2}\right),
\end{align}
where $\Gamma_i=E_1+\cdots+E_i$ and ${E_j, j \geq 1}$ are i.i.d. random variables with a standard exponential distribution.  To get the exact value of the innovations,  $K$ must tend to $\infty$ (here, we let $K=10,000$).    Perform this procedure again $n$ times independently to generate random numbers $\{\boldsymbol{\epsilon}_i\}$, $i=1,\ldots,n$.

\begin{figure}
         \centering
         \begin{subfigure}[b]{0.5\textwidth}
         \centering
                 \includegraphics[width=\textwidth]{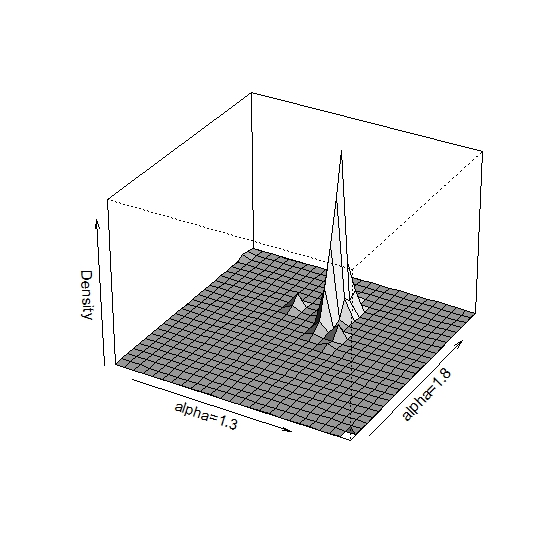}
                 \caption{$(\alpha_1=1.3,\alpha_2=1.8)$ }
                \label{fig:gull}
         \end{subfigure}%
         \begin{subfigure}[b]{0.5\textwidth}
         \centering
                 \includegraphics[width=\textwidth]{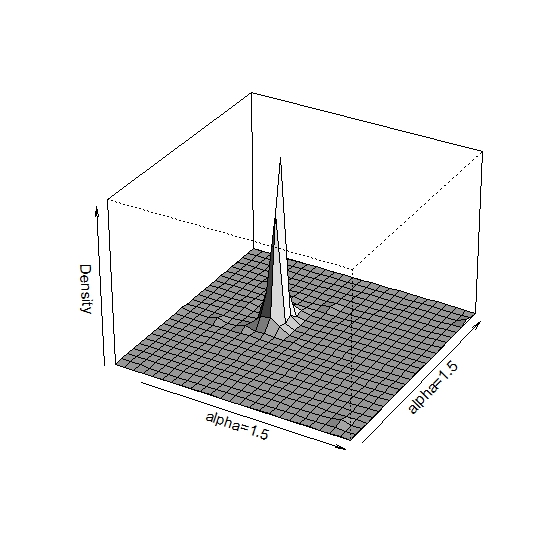}
                 \caption{$(\alpha_1=1.5,\alpha_2=1.5)$}
                 \label{fig:tiger}
         \end{subfigure}
\caption{Density  plot  for the bivariate observations $\left(X_{i1} , X_{i2}\right)$ given in \eqref{BiM}.}
         \label{fig:1318}
\end{figure}

 To acquire an intuitive feel for the bivariate observations with different indices of stability, we generate errors from \eqref{ME} with indices of stability
$(\alpha_1,\alpha_2)=(1.3,1.8)$ and $(1.5,1.5)$ and sample size $n=1,000$. The observations $\left(X_{i1} , X_{i2}\right)$, $i=1,\ldots,n$, are simulated  from \eqref{BiM}, and based on these observations, we plot  the joint density of $\left(X_{i1}, X_{i2}\right)$, $i=1,\ldots,n$.
Figure \ref{fig:1318} presents the  joint density of $\left(X_{i1} , X_{i2}\right)$  when $(\alpha_1,\alpha_2)=(1.3,1.8)$ and $(\alpha_1,\alpha_2)=(1.5,1.5)$, left to right respectively.

To illustrate the results of Theorem \ref{CONF INT}, we perform the following simulation study to construct $100(1-\alpha)\%$ confidence region for the mean vector in model \eqref{BiM}.  All the corresponding distributions of the innovations come from symmetric bivariate stable laws  with indices of stability
$(\alpha_1,\alpha_2)=(1.2,1.1)$, $(1.5,1.5)$, $(1.5,1.9)$, $(1.3,1.8)$, and $(2.0,1.2)$ with sample sizes $n=100$, $200$, and $500$. The simulation scheme for each choice of $n$ and $(\alpha_1,\alpha_2)$ is as follows:
\begin{enumerate}[(i)]
  		\item Generate $\{\boldsymbol{\epsilon}_{i}\}$ in \eqref{BiM} with the preceding indices of stability.
     \item Find $\{\boldsymbol{X}_{i}\}_{i=1}^{n}$ from \eqref{BiM}. Then estimate $\boldsymbol{\mu}$ by $\hat{\boldsymbol{\mu}}_{M}$ using the bivariate convex function $\rho$ given in \eqref{MultiHub} and   apply the Huber loss function in \eqref{HLF} for $\rho_1$ and $\rho_2$. Note that, according to the values of $\alpha_1$ and $\alpha_2$, the values of $c_1$ and $c_2$ in the Huber loss function are chosen from Table \ref{tab:TAlphaC} and plugging them in the result in Theorem \ref{CONF INT}.
     \item Estimate $\Sigma$ by replacing $\boldsymbol{\epsilon}_{i}$ with the residuals, $\boldsymbol{e}_{i}=\mathbf{X}_i-\hat{\boldsymbol{\mu}}_M, \text{ }i=1,2,\ldots,n$, in \eqref{Cov1} and \eqref{Cov2}.
     \item Estimate the naive $(1-\alpha)$-percentiles from $B=3,000$ bootstrap replications of $C^*=n(\hat{\boldsymbol{\mu}}_M^*-\hat{\boldsymbol{\mu}}_M)^T\mathbf{S}^{*^{-1}}(\hat{\boldsymbol{\mu}}_M^*-\hat{\boldsymbol{\mu}}_M)$. To do so, draw a sample of size $n$ from centered residuals denoted by  $\hat{\boldsymbol{\epsilon}}_1^*,\ldots,\hat{\boldsymbol{\epsilon}}_n^*$, and find  $\{\boldsymbol{X}_i^*\}_{i=1}^n$ from \eqref{BiM}. Then   estimate the parameters $\left(\mu_1 , \mu_2\right)$ using the bootstrap observations and  the same minimization technique in  step (ii). Use 3,000 bootstrap replications to compute the $(1-\alpha)$ quantile of $C^*(b),b=1,2,\ldots,3000$.
             \item Compute the confidence region from \eqref{ConfRGB}.
          \end{enumerate}

\begin{figure}[h!]
         \centering
         \includegraphics[width=0.5\textwidth]{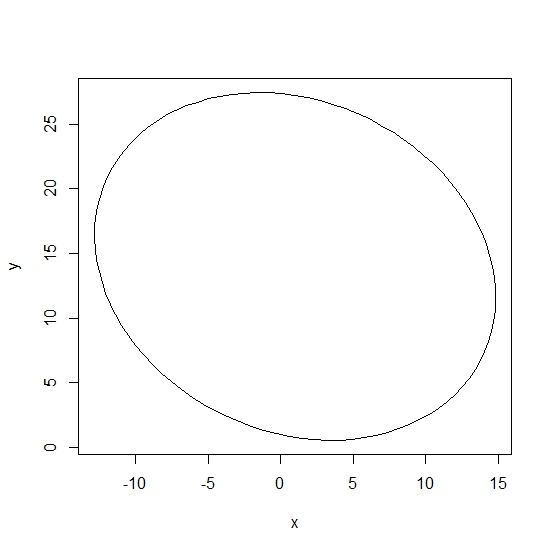}
\caption{Confidence region  for $\left(\mu_{1} , \mu_{2}\right)=\left(1 , 14\right)$ where $n=100$ and $(\alpha_1=1.5,\alpha_2=1.5)$ using the Huber loss function.}
         \label{fig:CR}
\end{figure}
Figure \ref{fig:CR} presents the confidence ellipse for a 95\% confidence level centered at the point  $\hat{\boldsymbol{\mu}}_{M}$ for sample size $n=100$ and $(\alpha_1=1.5,\alpha_2=1.5)$ by using the Huber loss function. This experiment was repeated $1,000$ times to estimate the coverage probabilities by checking whether  $\left(\mu_1 , \mu_2\right)=\left(1 ,14\right)$ are located within the estimated confidence region or not. Table \ref{tab:CPH} presents  the coverage probabilities for different combinations of $\alpha_1$ and $\alpha_2$  when we set confidence levels equal to $0.90$, $0.95$ and $0.99$ using the Huber loss function. As seen in Table \ref{tab:CPH}, the M-estimates  provide significantly  precise estimation such that the coverage probabilities for their related confidence region are fairly close to the nominal confidence levels.\\
\begin{table}[!h]
\centering
\caption{Estimated coverage probabilities by employing different choices of sample size, indices of stability, and confidence levels using the Huber loss function}
\begin{tabular}{lccccccccccccc}
  \hline
  {}  & {} & \multicolumn{3}{c}{$n=100$} &  {} &  \multicolumn{3}{c}{$n=200$} &  {} & \multicolumn{3}{c}{$n=500$} \\ \cline{3-5} \cline{7-9} \cline{11-13}
  $(\alpha_1,\alpha_2)$ &  {} & $I_{0.90}$ & $I_{0.95}$ & $I_{0.99}$ & {}
   &$I_{0.90}$ & $I_{0.95}$ & $I_{0.99}$ & {} &
    $I_{0.90}$ & $I_{0.95}$ & $I_{0.99}$\\
  \hline
  (1.2,1.1) & {} &  0.889 & 0.944 & 0.986 & {} &  0.872 & 0.946 & 0.982 &   {} & 0.900   &  0.955 & 0.989\\
  (1.5,1.5) & {} & 0.893  & 0.933  & 0.979 & {} & 0.89  & 0.941 &   0.986  &  {}  &  0.909   &  0.947 & 0.989\\
  (1.5,1.9) & {} &  0.889 & 0.934  & 0.985 & {} & 0.889  & 0.946  &  0.974 &  {}  & 0.887    & 0.952  & 0.988\\
  (1.3,1.8) & {} & 0.892 & 0.938 & 0.983 & {} &   0.900  &  0.943 & 0.991 & {}  & 0.896 & 0.953  & 0.987\\
  (2.0,1.2) & {} & 0.902 & 0.949 & 0.986  & {} & 0.901 & 0.958 & 0.987  & {} & 0.879  & 0.955  & 0.99\\
  \hline
\end{tabular}
\label{tab:CPH}
\end{table}
\bigskip

\section{Some notes and remarks}
In this paper, we address the problem of making robust inference about the population
mean when the observations are a sequence of  i.i.d.  random vectors   in the domain of attraction of  a stable law with possibly different indices of stability. In this case, we develop the M-estimate method that provides an efficient alternative to the sample mean. As we show in Theorem 1,  the M-estimate technique yields a central limit theorem  with the normal limiting distribution.

One of the most important  concerns in heavy-tailed phenomena is lack of knowledge about the tail indices.   Athreya et al. (1998)  consider the estimation of the population mean for a univariate case via the sample mean in which  the test statistic  depends on an unknown norming constant, $a_n$. They prove that, asymptotically, $a_n$ can be replaced by the maximum of the sample,$M_n$. This is due the fact that we do not know  the precise tail indices and we can only estimate them. However, to our knowledge, for the multivariate case, the estimation of the tail indices  is an open problem. On the other hand, the main advantage of  Theorem 1 is that the rate of convergence of the  M-estimate is $\sqrt{n}$ which is free from the tail indices $\alpha_i$, $i=1,\ldots,p$. Via the M-estimate method,  we only need to estimate  tail indices for  the tuning constants $c_j$'s in the Huber loss function.  However, Table 1  describes that the precise choice of the tuning constants $c_j$ for the specific values of the tail indices is not  critical in our estimation.  We still get an acceptable performance with a moderate value of $c_j$, i.e., a value close to the standard value
of 1.345 typically used in robust statistics.  Of course, some knowledge of indices of stability will improve our estimation.


The M-estimate method in this paper also rectifies the failure of the regular bootstrap as we discuss in Section 3. In practice, choosing a proper subsampling size to make the bootstrap consistent  is of great concern. Adopting this view, via the M-estimate method, the regular bootstrap is $\sqrt{n}$ consistent  with a Gaussian limit.


Notice that  it is not necessary to enforce the condition that $1<\alpha\leq2$ in model \ref{1.1}.  For $0<\alpha\leq1$, $E(\mathbf{X}_i)$, $i=1,\ldots,n$, does not exist. Therefore, $\boldsymbol{\mu}$ is not the mean for the observations $\{\mathbf{X}_i\}$ in \eqref{1.1} but we can still consider $\boldsymbol{\mu}$ as the shifting parameter. The estimation procedure and the asymptotic behaviour of our M-estimate remains valid and bootstrapping still works when we wish to estimate the shifting parameter $\boldsymbol{\mu}$.


 Liu,  Parelius, and  Singh (1999) consider a nonparametric multivariate method based on the concept of data depth to analyze multivariate distributional characteristics such as location, scale, bias, and skewness. However, this approach is not applicable if the moments do not exist. Similar extensions can be developed to the observations in the domain of attraction of a multivariate stable law. While it would  be of great interest to pursue these extensions, these approaches need to be further investigated in a separate study and will be dealt with in future research.

\bigskip
\noindent{\textbf{ Acknowledgments:}} This research is supported by a research grant from the \emph{Natural Sciences and Engineering Research Council of Canada (NSERC)}.

\end{document}